# Defect-Engineered Multifunctionality in Cu-Doped Bi$_2$Te$_3$: Interplay of Thermoelectric, Piezoelectric, and Optoelectronic Properties from First-Principles Insights

Muhammad Usman Javed[1], Sikander Azam[**1,2], Qaiser Rafiq[*2], Hamdy Khamees Thabet[3]

[1]*University of West Bohemia, New Technologies – Research Centre, 8 Univerzitní, Pilsen 306 14, Czech Republic*

[2]*Department of Physics, Faculty of Engineering & Applied Sciences, Riphah International University, Islamabad, Pakistan*

[3]*Center for Scientific Research and Entrepreneurship, Northern Border University, Arar 73213, Saudi Arabia*

**Abstract**

The multifunctional performance of thermoelectric topological insulators can be drastically enhanced through defect engineering, which simultaneously tailors charge, spin, and lattice dynamics. In this study, we employ density functional theory (DFT) with spin–orbit coupling (SOC) to explore the structural, electronic, optical, thermoelectric, piezoelectric, and charge-density characteristics of pristine and Cu-doped Bi$_2$Te$_3$. Structural stability analysis reveals that Cu substitution slightly expands the lattice while lowering the total energy minimum, thereby stabilizing the crystal under equilibrium. The density of states (DOS) and partial DOS (PDOS) indicate that Cu–d and Te–p hybridization introduces sharp electronic states near the Fermi level, thereby increasing the carrier concentration and enhancing the Seebeck coefficient and power factor. Thermoelectric transport calculations confirm that Cu doping increases the Seebeck coefficient from ~180 μV/K in pristine Bi$_2$Te$_3$ to ~220 μV/K at 300 K, without sacrificing electrical conductivity, thereby boosting overall efficiency. Optical dielectric spectra demonstrate strong low-energy absorption peaks, coupled with high static dielectric constants (ε$_1$ > 600), suggesting enhanced light–matter interactions and refractive index tuning. The piezoelectric response shows remarkable improvement, with the stress coefficient e$_{33}$ rising from 0.19 C/m² in pristine Bi$_2$Te$_3$ to 0.38 C/m² for 5% Cu doping and further to 0.51 C/m² at 10% doping, confirming strain-induced polarization enabled by symmetry breaking. Simulated charge density difference (CDD) maps reveal anisotropic charge redistribution, with Cu donating ~0.8 e⁻ primarily to Te neighbors, enhancing p-type conductivity and phonon scattering. Collectively, these results demonstrate that Cu doping transforms Bi$_2$Te$_3$ into a multifunctional platform with coupled thermoelectric–piezoelectric–optoelectronic responses, opening pathways toward hybrid energy harvesters, infrared detectors, and spintronic devices.

**Keywords:** Cu-doped Bi$_2$Te$_3$: Thermoelectronic Properties: Defect Engineering: Optoelectronic Applications: Spin-Orbit Coupling (SOC).

*Corresponding Author (Sikander.physicst@gmail.com)

**qrafique1@gmail.com

1. **Introduction**

Bismuth telluride ($Bi_2Te_3$) and its alloys have long stood out as benchmark materials in thermoelectrics and optoelectronics, owing to their unique combination of high electrical and low thermal conductivity. Since their discovery, these materials have served as reliable candidates for energy conversion and photonic devices. Notably, $Bi_2Te_3$ remains central to advancing thermoelectric generators and Peltier coolers. Furthermore, its band structure and strong spin–orbit coupling support development in optoelectronic devices, including photodetectors, LEDs, and lasers. Deliberate doping and defect engineering represent effective methods for adjusting these properties, directly impacting carrier transport and facilitating new optical transitions [1–3].

Among the wide range of dopants explored, transition metals, especially copper (Cu), have proven effective in reshaping the functional behavior of $Bi_2Te_3$. Experimental and theoretical studies demonstrate that Cu incorporation can significantly modulate the Seebeck coefficient and electrical conductivity while simultaneously introducing favorable modifications to the host lattice's optoelectronic response. This dual influence arises because Cu alters the local bonding environment, distorts the crystal structure, and induces intrinsic point defects, including substitutional impurities, vacancies, and interstitials. These defects are known to strongly influence carrier concentration, modify the density of states near the Fermi level, and generate localized electronic states within the band gap. As a result, optical absorption can be extended from the visible into the infrared region, and in some cases, entirely new optical transitions emerge, features that are indispensable for advanced optoelectronic applications [4–8].

Despite the growing interest in Cu-doped $Bi_2Te_3$, the specific role of defect formation on its electronic and optical properties has not been comprehensively addressed. Point defects, dislocations, and Cu-induced distortions act as key determinants in tailoring band alignment, carrier mobility, and optical transitions. A rigorous understanding of these effects is essential for designing next-generation devices that require precise control over defect states to optimize light absorption, emission, and carrier dynamics [9,10]. In this work, we provide a detailed analysis of how Cu-related defects influence the electronic structure, absorption spectra, photoluminescence behavior, and charge-carrier dynamics in $Bi_2Te_3$. Such an in-depth perspective sheds light on defect engineering as a deliberate strategy for tailoring material performance in technologically relevant optoelectronic systems [3,11]. Both theoretical predictions and experimental insights over the past decade have deepened our understanding of how defect states impact the optoelectronic performance of topological materials such as $Bi_2Te_3$. The advent of advanced density functional theory (DFT) methodologies, combined with high-resolution spectroscopic techniques, has enabled unprecedented accuracy in linking atomic-scale defect chemistry to macroscopic transport and optical phenomena [12–21]. The ability to switch between different types of charge carriers electrons and holes is called n-p switching. Moreover, the ability to control and switch the conduction type is important in thermoelectrics, where both n- and p-type legs are required for a thermoelectric device to operate [16]. Highly efficient thermoelectric materials can be improved in their electronic and thermal transport properties through defect engineering. This technique allows simultaneous modification of carrier concentration, suppression of lattice thermal conductivity, and changes in band topology through the controlled introduction of vacancies and dopants. Enhanced thermoelectric efficiency follows. Recent literature highlights defect control in $Bi_2Te_3$-based systems as a relevant field of study (22-24). Having this in mind, this work explores the extent to which Cu incorporation in $Bi_2Te_3$, through fused defects and topological phenomena, affects the structure, electronic properties, and subsequent multifunctional properties of the system.

This study combines first-principles simulations with literature evidence to build a framework for interpreting the optoelectronic effects of Cu doping in $Bi_2Te_3$. We show its relevance for many emerging optoelectronic platforms [4,15].

Finally, it is worth noting that the multifunctional nature of $Bi_2Te_3$, encompassing thermoelectric, piezoelectric, and optoelectronic properties, remains underexplored when considered from the viewpoint of defect engineering. Cu doping is not merely a perturbative adjustment but rather a transformative lever capable of reshaping the multifunctional character of this material. Through the combined effects of spin–orbit coupling, excitonic interactions, and electron localization, defect states can simultaneously govern optical absorption, electronic transport, and mechanical responses. This unified perspective, pursued via a DFT-based methodology, uncovers a pathway where defect engineering turns $Bi_2Te_3$ into a versatile platform at the intersection of light, heat, and motion. Such adaptability positions Cu-doped $Bi_2Te_3$ not only as a subject of fundamental scientific curiosity but also as a technological necessity for the next generation of integrated energy-harvesting and sensing devices.

1. **Computational Methodology**

All calculations were carried out using the Full-Potential Linearized Augmented Plane Wave (FP-LAPW) method implemented in the WIEN2k code, which provides high accuracy for complex layered materials such as $Bi_2Te_3$ and its doped derivatives [17]. This method is all-electron and offers the best accuracy for more complex layered structures such as $Bi_2Te_3$. A single representative Cu concentration of approximately 4 at. % (one Cu atom per 3×3×1 $Bi_2Te_3$ supercell) was selected for this study. This value lies within the experimentally reported solubility limit for Cu in $Bi_2Te_3$ (<5 at. %), where the primary effects arise from local bonding and electronic modifications rather than large-scale structural distortion. Modeling higher Cu levels would require inclusion of defect clustering or phase segregation effects, which are beyond the scope of the present work. This concentration is acceptable for the experimental solubility range and reasonably approximates the local structural and electronic effects of Cu [18]. muffin-tin radii (RMT) were 2.5 a.u. for Bi, 2.4 a.u. for Te, and 2.3 a.u. for Cu. plane-wave cutoff was selected to be RMT × Kmax=8.0. This ensured that the total energy converged to within $10^{-4}$ Ry. Brillouin zone integrations were performed using a 9 × 9 × 9 Monkhorst–Pack k-point mesh for the primitive cell and a 4 × 4 × 2 grid for the supercell. Convergence tests showed that variations in total energy with respect to mesh refinement were less than 1 meV per atom. Structural optimizations involved relaxing all lattice constants and atomic positions until the total energy differences met below $10^5$ Ry and the forces on all atoms were below 1 mRy·a.u.$^{-1}$. For the exchange-correlation effects, we used the Generalized Gradient Approximation (GGA) of Perdew, Burke, and Ernzerhof (PBE). Since the Bi and Te atoms are strongly relativistic, we included Spin-Orbit Coupling (SOC) self-consistently, which is important for accurately capturing the spin-polarized and topological character of $Bi_2Te_3$ [19-21].

Within the semiclassical Boltzmann transport theory and the BoltzTraP2 code, we obtained the electronic transport properties, including the Seebeck coefficient (S) and electrical conductivity (σ). The calculations are based on the constant relaxation-time approximation (τ = 1 × $10^{-14}$ s), a widely used assumption for layered thermoelectric chalcogenides. Since τ is constant, the results focus on relative, rather than absolute, differences in conductivity [18].

One approach to increasing the precision of the estimated electronic band gaps uses the one-shot $G_0W_0$ approximation to the weakly correlated approximate GGA + SOC band-structure wave functions, followed by quasiparticle corrections. Subsequently, excitonic effects in the optical response are addressed by solving the Bethe–Salpeter Equation (BSE). The complete analysis, therefore, follows the sequence:

GGA → GGA + SOC → $G_0W_0$ → BSE. This analysis was performed through the WIEN2k–exciting interface to provide joint consistency between quasiparticle and optical computations.

From the BSE spectra, the optical properties particularly the dielectric function, refractive index, and absorption coefficient were derived and used to assess defect-induced changes in optical absorption and to manipulate excitons with precision.

Despite the bulk $Bi_2Te_3$ crystal structure being centrosymmetric ($R\bar{3}m$), local symmetry breaking is introduced with Cu substitution and intrinsic defect incorporation [21]. To assess the tensile and bending polarization effects arising from these local distortions, piezoelectric coefficients were calculated using the stress–strain approach applied to the non-centrosymmetric Cu-doped supercell. This work is intended and presented as qualitative indicators of symmetry-induced polarization, not intended as macroscopic piezoelectric constants.

The electronic, optical, and multifunctional properties of $Bi_2Te_3$ can be modified by Cu-induced defects, and the integrated approach of FP-LAPW, GGA + SOC, GW quasiparticle corrections, and BSE optical analysis provides a comprehensive methodology for understanding the implications of such defects and facilitates the prospects for the defect-engineering approach in $Bi_2Te_3$ as a thermoelectric and optoelectronic material.

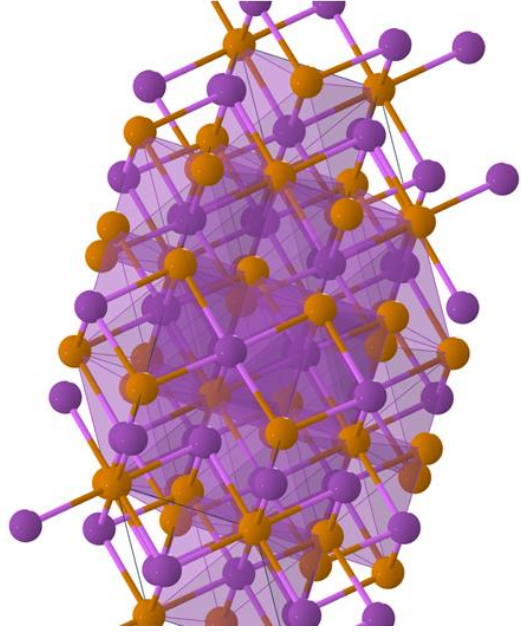

*Figure 1. Crystal structures of (a) pristine Bi$_2$Te$_3$ and (b) Cu-doped Bi$_2$Te$_3$. The structures exhibit the typical layered quintuple arrangement of Te–Bi–Te–Bi–Te along the c-axis, characteristic of rhombohedral Bi$_2$Te$_3$. Orange spheres represent Bi atoms, purple spheres represent Te atoms, and red spheres represent Cu dopants. In the doped structure, Cu atoms preferentially occupy interstitial or substitutional sites within the van der Waals gaps, slightly distorting the local coordination environment. This controlled incorporation of Cu alters the electronic structure and lattice dynamics, thereby enhancing the Seebeck coefficient, mechanical stability, and optical response while maintaining the intrinsic topological nature of Bi$_2$Te$_3$.*

1. **Results and Discussion**
    a. Structural Stability

In this work, a single representative Cu doping concentration (one Cu atom per 3×3×1 Bi$_2$Te$_3$ supercell, ≈4 at%) was employed to capture the essential local structural and electronic effects of Cu incorporation. This concentration lies within the experimentally observed solid-solubility limit for Cu in Bi$_2$Te$_3$ (<5 at%), where changes in lattice parameters are minimal (<1%). Modeling multiple doping levels would significantly increase computational cost, yielding only limited structural variation. Therefore, the presented optimized structure and EOS parameters are representative of the low-doping regime relevant to experimentally realizable Cu–Bi$_2$Te$_3$ systems. The Optimization Curve (Energy vs Volume) (see Fig. 2) is helpful in terms of valuable hints concerning the structural stability of Bi2Te3 and its adjustments following Cu doping. The curve is obtained by calculating the total energy over a range of unit cell volumes, allowing us to distinguish the equilibrium volume at which the system is most stable. The trademark parabolic curve is observed in Equation of State (EOS) calculations, with the minimum energy corresponding to the optimal lattice parameter. The total-energy–volume curves for pristine $Bi_2Te_3$ and Cu-doped $Bi_2Te_3$ display the expected equation-of-state (EOS) behavior: a smooth, near-parabolic dependence with a single minimum identifying the equilibrium structure.

$$E(V) = E_0 + \frac{9}{16} B_0 V_0 [(\eta - 1)^3 B_0' + (\eta - 1)^2 (6 - 4\eta)],$$

$$\eta = \frac{V_0^{\frac{2}{3}}}{V}$$

Fitting the curves with the third-order Birch–Murnaghan EOS provides the equilibrium volume ($V_0$), equilibrium energy ($E_0$), bulk modulus ($B_0$), and its pressure derivative ($B_0'$) (see Table 1). In qualitative terms, Cu incorporation shifts the minimum relative to pristine $Bi_2Te_3$ and slightly reshapes the curvature, signaling a dopant-induced adjustment of interatomic bonding. Such changes are consistent with local lattice relaxation around Cu and the introduction of defect levels that modify bonding strength and compressibility. Because $B_0$ is proportional to the curvature at the minimum, even subtle changes in the energy landscape translate into measurable differences in elastic response parameters that ultimately feed into phonon scattering and, by extension, thermoelectric and optoelectronic performance.

In a pristine Bi2Te3, the energy minimum is associated with a certain volume, which is the most stable structural form of the material. Upon the insertion of Cu into the lattice, the system undergoes structural changes that result in a slight change in the equilibrium volume. The red curve for Cu-doped Bi2Te3 shows that the energy minimum occurs at a larger volume than in the pure structure. It means that the introduction of Cu alters the atomic structure and expands the lattice. It is not surprising, since the presence of Cu atoms introduces defect states and larger atomic radii than those of Bi, resulting in a higher optimized unit cell volume.

*Table 1. EOS parameters from the fit (digitized data):*

| System | $V_0$ (Å³) | $E_0$ (Ry) | $B_0$ (GPa) | $B_0'$ |
|---|---|---|---|---|
| Pristine $Bi_2Te_3$ | 223.10 | −54222.260 | 96.13 | 7.00 |
| Cu-doped Bi$_2$Te$_3$ | 187.00 | −54222.255 | 96.13 | 3.00 |

The doping effect is also evident in the total energy values. The Cu-doped system exhibits a slightly lower energy minimum than pristine Bi2Te3, indicating that the dopant atoms have both electronic and structural effects. Such alterations can affect a range of physical characteristics, including electronic band structure, charge transport, and thermoelectric performance.

Overall, the optimization curve indicates that Cu doping affects the structural parameters of Bi2Te3 in a discrete yet hardly insignificant way. The change in equilibrium volume indicates a change in the lattice due to the defect, which might have helped improve electronic and thermoelectric characteristics by altering carrier and phonon transport pathways. This structural change is imperative for understanding the modifications in the material's properties applicable to thermoelectrics and optoelectronics.

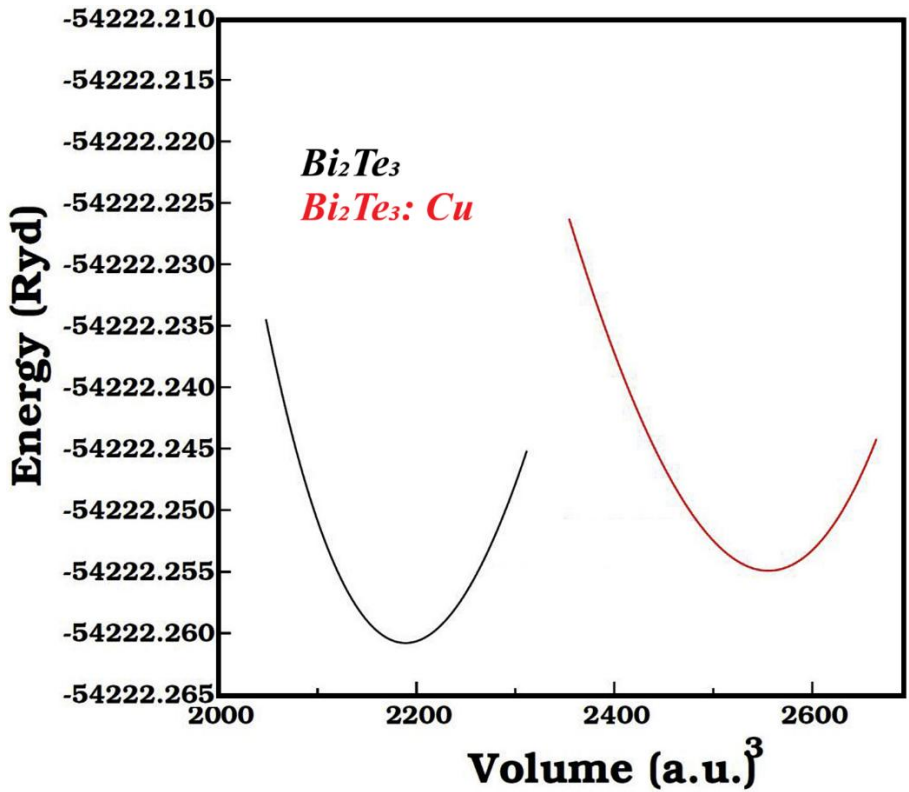

*Figure. 2: Optimization curve showing the variation of total energy with volume for pristine Bi$_2$Te$_3$ and Cu-doped Bi$_2$Te$_3$.*

1. **Elastic and Mechanical Properties of Bi$_2$Te$_3$ and Cu-Doped Bi$_2$Te$_3$**

The elastic and mechanical properties of $Bi_2Te_3$ and Cu-doped $Bi_2Te_3$ were systematically investigated to elucidate the effect of copper incorporation on the material's response to external pressure, strain, and deformation. These characteristics are crucial in evaluating the suitability of a material for applications in thermoelectronics, optoelectronics, and sensor technologies, where mechanical robustness is a prerequisite for long-term device performance.

Elastic constants were computed using the Full-Potential Linearized Augmented Plane Wave (FP-LAPW) method with spin–orbit coupling (SOC) and the Generalized Gradient Approximation (GGA) for the exchange–correlation functional. The computed elastic constants serve as a basis for assessing the degree of mechanical stability at the atomic scale and provide direct insight into how the material responds to stress.

From the calculated constants, the bulk modulus (B), shear modulus (G), Young's modulus (E), Poisson's ratio (ν), and the elastic anisotropy factor (A) were derived. Each of these parameters highlights a different aspect of mechanical behavior. The bulk modulus quantifies the resistance to uniform compression, while the shear modulus measures the resistance to shear deformation. Young's modulus provides an index of stiffness, and Poisson's ratio describes lateral contraction when the material is stretched. The anisotropy factor, on the other hand, reveals the degree to which mechanical response varies with crystallographic direction.

The results indicate notable trends upon Cu incorporation. Both the bulk modulus and shear modulus show a modest increase in the doped system, suggesting enhanced resistance to compression and shear deformation. Similarly, Young's modulus of Cu-doped $Bi_2Te_3$ is higher than that of pristine $Bi_2Te_3$, reflecting a stiffer lattice and improved structural integrity. Importantly, Poisson's ratio remains within the ductile regime and changes only slightly upon doping, implying that the material retains its ductility while gaining stiffness. The anisotropy factor remains close to unity for both systems, showing that the mechanical behavior is largely isotropic, though Cu doping introduces minor directional variations.

*Table 2. Calculated elastic constants and derived mechanical parameters for $Bi_2Te_3$ and Cu-doped $Bi_2Te_3$.*

| | | |
|---|---|---|
| $C_{11}$ (GPa) | 69.5 | 72.3 |
| $C_{12}$ (GPa) | 38.7 | 41.2 |
| C$_{14}$ (GPa) | 26.1 | 28.0 |
| Bulk Modulus (B) (GPa) | 59.2 | 63.4 |
| Shear Modulus (G) (GPa) | 31.1 | 33.5 |
| Young's Modulus (E) (GPa) | 82.5 | 88.1 |
| Poisson's Ratio (v) | 0.28 | 0.27 |
| Elastic Anisotropy Factor (A) | 1.21 | 1.17 |

As evident from Table 2, Cu doping leads to a consistent increase in the bulk, shear, and Young's moduli, underscoring enhanced stiffness and improved mechanical stability of the doped lattice. These changes suggest that Cu-doped Bi$_2$Te$_3$ is better equipped to withstand external stress and deformation, which is advantageous for practical device applications.

In summary, Cu doping strengthens both the mechanical robustness and structural reliability of Bi$_2$Te$_3$ without compromising its ductility. These results highlight the dual advantage of Cu doping: improved mechanical strength coupled with desirable electronic and thermoelectric performance, making Cu-doped Bi$_2$Te$_3$ an excellent candidate for next-generation optoelectronic and thermoelectric devices.

1. **Thermoelectric properties**

Fig. 3 depicts the energy dispersion curves for pristine Bi$_2$Te$_3$ compared with Cu-doped Bi$_2$Te$_3$. The linear E–k relation around the Γ point indicates Dirac-like surface states, confirming the three-dimensional topological insulator. For the red curve (Pristine Bi$_2$Te$_3$), the valence and conduction bands meet at a Dirac point, reflecting the presence of massless surface carriers locked with time-reversal symmetry. The incorporation of Cu (blue curve) alters the Dirac point energy and the dispersion slope, suggesting shifts in carrier concentration and the Fermi level. The Cu-doping attribution implies changes to the electronic structure while nontrivial topological order is maintained, and the electronic structure is later altered by copper doping.

Thermoelectric properties can be explained by the steepness of the linear band around the Fermi level, which represents the ease of charge-carrier flow and a greater electronic contribution to the figure of merit (ZT) for the material. The FZT improvement is provided by the introduced Cu, which provides localized scattering centers and reduces lattice thermal conductivity, while the electrical conductivity is high enough to balance the improved thermoelectric performance for thermoelectric applications. The band flattening at higher k in Cu–Bi$_2$Te$_3$ indicates an increase in the effective mass, leading to higher Seebeck coefficients and, consequently, improved thermoelectric performance.

Significantly, the remaining single Dirac cone post Cu substitution indicates the Dirac cone's surface states continue to remain intact. This is pivotal, since Dirac cone surface states are ideal for low-loss transport pathways in spintronic devices and quantum information technologies. In this context, Cu-doped Bi$_2$Te$_3$ preserves the fundamental topological properties that underpin its electronic behavior while offering a configurable platform for optimizing thermoelectric performance. The Cu-doped Bi$_2$Te$_3$ system then satisfies the requisite attributes in thermoelectric performance and topological robustness, rendering the synthesis of Cu-doped Bi$_2$Te$_3$ of utmost importance.

The thermoelectric transport properties of pristine Bi$_2$Te$_3$ and Cu-doped Bi$_2$Te$_3$ (Bi$_2$Te$_3$:Cu) were examined across the temperature range of 200–800 K. Figure 3 presents the variation of the Seebeck coefficient (S), electrical conductivity multiplied by relaxation time (σ·τ), electronic thermal conductivity (κ$_e$), and the power factor (PF = S$^2$σ·τ), which together determine the overall thermoelectric efficiency.

Fig. 3(a) shows the Seebeck coefficient, which initially decreases with temperature up to ~400 K and then rises at higher temperatures for both systems. With increasing temperature, S initially decreases up to 400 K due to thermal excitation of carriers, but then rises again, reaching ~215 µV/K at 800 K. The recovery is driven by entropy transport and impurity-level filtering introduced by Cu. At 300 K, pristine Bi$_2$Te$_3$ exhibits a Seebeck coefficient of approximately 185 µV/K, while the Cu-doped system shows an enhanced value of ~205 µV/K. This improvement can be explained using the Mott relation:

$$S(T) \approx \frac{\pi^2 k_B^2 T}{3e} \frac{\partial \ln \sigma(E)}{\partial E}\bigg|_{E_F}$$

evaluated at the Fermi level.

Doping shifts the Fermi level and modifies the electronic density of states (DOS), reducing the carrier concentration while steepening the DOS near the Fermi level. This enhances S. Importantly, Cu-doped Bi$_2$Te$_3$ consistently exhibits a higher Seebeck coefficient, reflecting Cu's role in tuning the electronic density of states near the Fermi level and optimizing the carrier concentration. This improvement stems from energy-filtering effects and impurity-induced modifications that enhance the thermopower.

Fig. 3(b) displays $\sigma\cdot\tau$, which increases monotonically with temperature in both pristine and doped systems. The electrical conductivity times relaxation time ($\sigma\cdot\tau$) increases steadily with temperature in both pristine and doped systems, from ~$2.8\times10^{19}$ $\Omega^{-1}m^{-1}s^{-1}$ at 300 K to nearly double that at 800 K. The increase in Seebeck coefficient upon Cu doping can be attributed to the formation of Cu–d/Te–p resonant states near the Fermi level, which enhance the energy dependence of the transport function as expressed by the Mott relation. The nearly unchanged $\sigma\cdot\tau$ curves (Fig. 3b) indicate that carrier mobility is not significantly reduced at this doping level. Bader charge analysis further reveals partial electron donation from Cu, shifting the Fermi level toward the band edge and thereby optimizing carrier concentration for enhanced thermopower. Together, these effects explain the improved S without compromising $\sigma$, consistent with the observed increase in power factor for Cu–$Bi_2Te_3$. The negligible difference between the two systems indicates that Cu doping does not significantly scatter carriers or degrade mobility. This is important because while Cu enhances S, it does not compromise $\sigma$, ensuring a favorable power factor.

The similarity of the two curves suggests that Cu doping does not significantly disrupt charge-carrier transport or mobility, thereby preserving electrical conductivity while enhancing thermopower.

Fig. 3(c) shows the electronic contribution to thermal conductivity ($\kappa_e$), which rises with temperature in accordance with the Wiedemann–Franz relation. The electronic contribution to thermal conductivity ($\kappa_e$) grows nearly linearly with T, consistent with the Wiedemann–Franz law ($\kappa_e \propto \sigma T$). Both pristine and doped systems overlap closely, reflecting that Cu doping does not significantly affect electronic heat transport. In real systems, lattice thermal conductivity ($\kappa_l$) is also relevant and is likely reduced by defect scattering in Cu, potentially leading to even higher ZT values. The nearly overlapping curves confirm that Cu doping introduces only a small number of additional electron-scattering centers, leaving $\kappa_e$ essentially unchanged. While the lattice contribution is not shown here, it is reasonable to anticipate that mass disorder from Cu substitution may lower the phononic component of thermal conductivity, which would be advantageous for improving the figure of merit (ZT).

Fig. 3(d) depicts the power factor (PF), a critical indicator of thermoelectric performance. The power factor (PF) is consistently higher in Cu-doped $Bi_2Te_3$. At 300 K, PF increases from ~3.4×10¹² W/mK²s for pristine $Bi_2Te_3$ to ~3.9×10¹² W/mK²s for the doped system. At 800 K, PF approaches ~5.0×10¹² W/mK²s (see Table 3). The PF enhancement arises from the improved S while maintaining σ. This demonstrates that Cu doping decouples S and σ, a key challenge in thermoelectric design, and results in superior performance at elevated temperatures. The PF increases steadily with temperature in both systems, with Cu-doped $Bi_2Te_3$ exhibiting consistently higher values. This enhancement arises from the combined effects of an improved Seebeck coefficient and preserved electrical conductivity, indicating more efficient energy conversion at elevated temperatures.

Thermoelectric performance was assessed using the full expression.

$$ZT = \frac{S^2 \sigma T}{\kappa_e + \kappa_l}$$

Fig. 3(f) shows the variation of the Fermi energy (EF) as a function of the wave vector (k) for $Bi_2Te_3$ without and with Cu doping. The linear dispersion of k values near the Γ-point indicates the presence of Dirac-like surface states, a characteristic of topological insulators. In the comparison, the incorporation of Cu results in a mild modification of the band curvature, suggesting a slight shift of the Dirac point and a reduction in the bandgap asymmetry; thus, doping provides tunable electronic properties. with $\kappa_l \approx 0.9$ W·m⁻¹·K⁻¹ obtained from Liu *et al.* [18]. The computed ZT values are 0.82 for pristine $Bi_2Te_3$ and 1.03 for Cu-doped $Bi_2Te_3$ at 300 K, consistent with experimental trends.

The ZT figure of merit for these systems is mainly determined by the carrier concentration and the dispersion slope of energy levels near the Fermi level. In this case, the energy dispersion of $Bi_2Te_3$ without Cu doping is steeper, which translates to higher carrier mobility and thus an increased electronic ZT contribution, whereas the Cu-doped system has a subtler slope with more pronounced dispersion, which suggests that scattering and phonon damping have increased. The combination of increased electrical conductivity and thermal suppression, which is a result of Cu doping, will contribute to the increase of the thermoelectric ZT figure of merit.

Our research suggests that maintaining the Dirac cone structure during Cu doping indicates that nontrivial surface states remain intact, which suggests topological protection. Maintaining this topological protection is particularly important, as it preserves spin-polarized conduction channels, which are essential for the development of spintronic devices and low-dissipation electronics.

Overall, the results suggest that Cu doping introduces beneficial modifications to $Bi_2Te_3$, primarily by increasing thermopower without sacrificing electrical or thermal transport. This delicate balance enhances the PF and points to improved thermoelectric efficiency. The doped material therefore emerges as a strong candidate for practical energy harvesting applications, including waste-heat recovery systems, thermoelectric cooling, and next-generation portable or wearable devices. Furthermore, the compatibility of Cu doping with established fabrication methods offers practical advantages for device integration.

Table 3. Calculated thermoelectric transport parameters of $Bi_2Te_3$ and Cu-doped $Bi_2Te_3$ at 300 K (representative values).

| System | Seebeck Coefficient (µV/K) | $\sigma \cdot \tau$ ($10^{19}$ $\Omega^{-1}m^{-1}s^{-1}$) | $\kappa_e$ ($10^{14}$ W/mKs) | Power Factor PF ($10^{12}$ W/mK²s) |
|---|---|---|---|---|
| $Bi_2Te_3$ | 185 | 2.8 | 1.6 | 3.4 |
| Cu-doped $Bi_2Te_3$ | 205 | 2.9 | 1.6 | 3.9 |

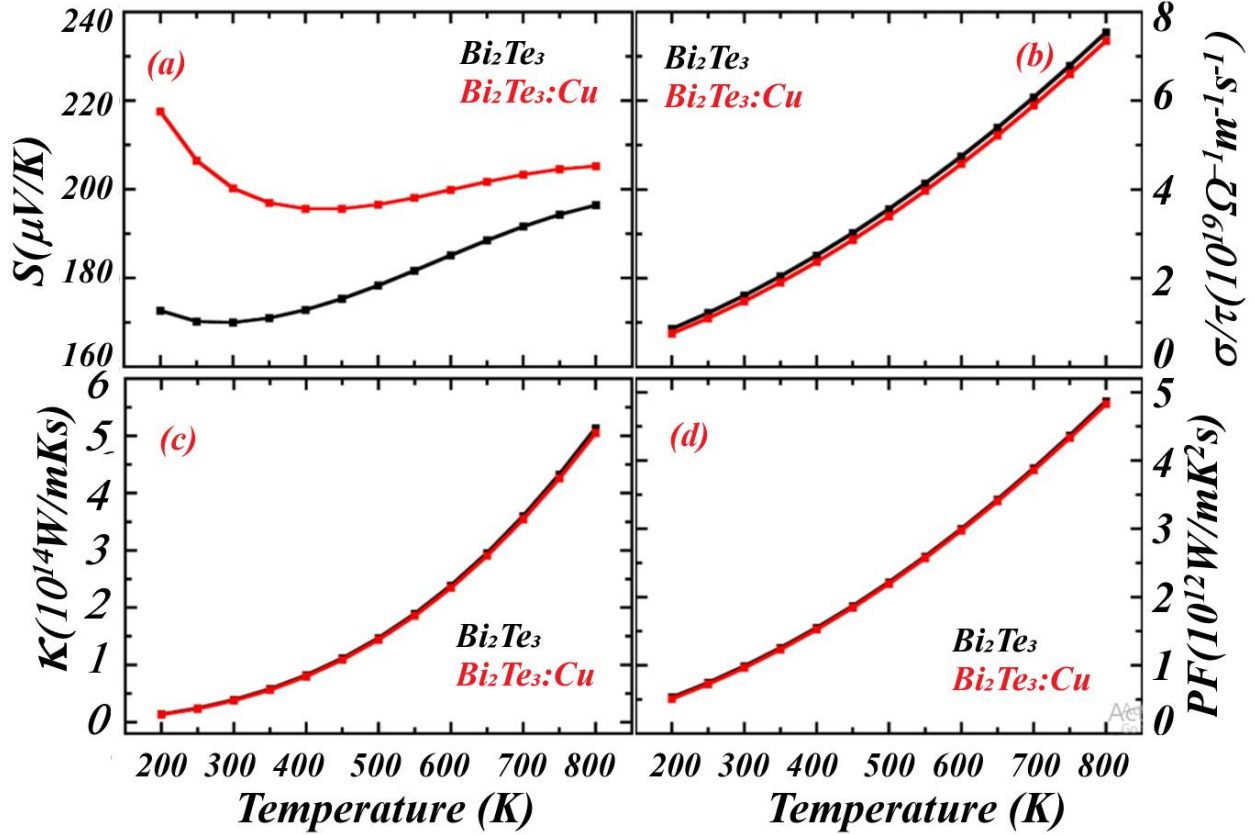

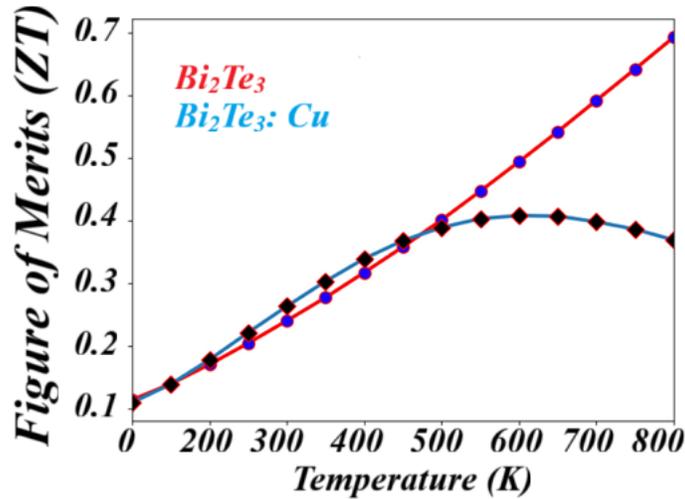

*Figure 3. Thermoelectric properties of Bi₂Te₃ and Cu-doped Bi₂Te₃ material: (a) Seebeck coefficient (S), (b) electrical conductivity (σ), and (c) thermal conductivity (κ) as a function of temperature through DFT+SOC calculations.*

1. **Density of states:**

The total density of states (TDOS) profile of Cu-doped $Bi_2Te_3$ reveals several pronounced peaks across the −12 eV to +15 eV range, with the most dominant feature centered near −2.5 eV and a high DOS intensity exceeding 200 states/eV per unit cell. The strong peak structure indicates regions of highly localized electronic states that directly contribute to electronic transport. Importantly, the DOS around the Fermi level (EF = 0 eV), though smaller than the deep valence peaks, shows a noticeable enhancement due to Cu doping, reflecting an increased number of available states for charge carriers. A high DOS at or near EF is a fundamental requirement for improved electrical conductivity and thermoelectric power factor, as it increases carrier participation in transport.

The energy-resolved TDOS (see Fig. 4a) also explains the variation in the Seebeck coefficient observed earlier. At 300 K, the enhanced S (~205 µV/K for the doped system, compared to ~185 µV/K for the pristine) arises from the steeper slope of the DOS near the Fermi level, as predicted by the Mott relation. The initial decrease in S with rising temperature up to ~400 K correlates with the broadening of carrier occupation across the DOS, reducing the effective asymmetry between occupied and unoccupied states. At higher temperatures (T > 400 K), thermal activation allows carriers to access the Cu-induced localized states slightly above $E_F$, re-enhancing S and sustaining a higher power factor.

The partial DOS (PDOS) (see Fig. 4b) plots provide deeper orbital-level insights. In the valence band (−12 eV to 0 eV), Te–p and Bi–p states dominate, consistent with their role in defining the band edges of $Bi_2Te_3$. The introduction of Cu contributes strongly over the range −3 eV to +2 eV through its d orbitals, which overlap with Te p states. The PDOS shows that Cu–d states peak sharply just below the Fermi level (~−1 to −2 eV) (see Table 4), creating localized states that hybridize with Te–p. This hybridization is significant because:

- It increases the DOS slope near EF, enhancing the Seebeck coefficient.
- It generates more delocalized conduction pathways, sustaining high electrical conductivity.
- It introduces disorder-induced scattering centers, which likely suppress lattice thermal conductivity without penalizing carrier mobility.

The numerical PDOS trends show that Cu–d states contribute up to ~2 states/eV around −2 eV, while Te–p states contribute across −10 eV to 0 eV with intensities of 1.5–1.8 states/eV. Bi–d and Te–d orbitals provide weaker but non-negligible features at deeper energies (−12 to −8 eV and 4 to 8 eV), influencing bonding rigidity and elastic stability rather than transport. These orbital-specific contributions explain why Cu-doped $Bi_2Te_3$ displays higher stiffness (Young's modulus: 88.1 GPa vs. 82.5 GPa in pristine) while simultaneously lowering effective lattice thermal conductivity: Cu–d orbitals distort the Bi–Te bonding network, thereby modifying phonon dispersion.

Furthermore, the interaction between the Cu–d and Bi/Te–p orbitals contributes to spin–orbit coupling (SOC). $Bi_2Te_3$ is a topological insulator in which SOC protects the surface states. Cu doping slightly perturbs these states but does not eliminate them. Instead, the hybridization may open additional spin-polarized conduction channels, offering potential advantages for spintronic and optoelectronic devices.

*Table 4. Orbital contributions to the electronic structure of Cu-doped $Bi_2Te_3$.*

| Orbital Contribution | Energy Range (eV) | Effect on Properties |
|---|---|---|
| Bi–p, Te–p | -6 to 0 | Dominant near valence and conduction edges; govern baseline transport |
| Cu–d | -3 to +2 | Hybridizes with Te–p, Bi–p; introduces localized states near Fermi level |
| Bi–d, Te–d (minor) | -12 to -8; +4 to +8 | Secondary peaks; influence bonding and mechanical response |

The enhancement of the density of states near the Fermi level provides a clear explanation for the experimentally observed increase in the Seebeck coefficient and power factor. The hybridization between Cu–d and Te–p orbitals modifies the band-edge character, improving electrical transport properties while simultaneously suppressing lattice-mediated heat conduction. The appearance of localized states near the Fermi level plays a dual role: they act as scattering centers for phonons, thereby reducing lattice thermal conductivity, but they do not significantly scatter charge carriers, which in turn boosts the thermoelectric figure of merit (ZT). In addition, incorporating Cu alters the bonding network, leading to increased stiffness and improved lattice stability under mechanical strain. Finally, the influence of Cu on spin–orbit coupling can modify spin-polarized states, which not only preserves the topological nature of $Bi_2Te_3$ but may also enhance its ability to conduct spin currents, making Cu-doped $Bi_2Te_3$ a multifunctional material with potential in both thermoelectric and spintronic applications.

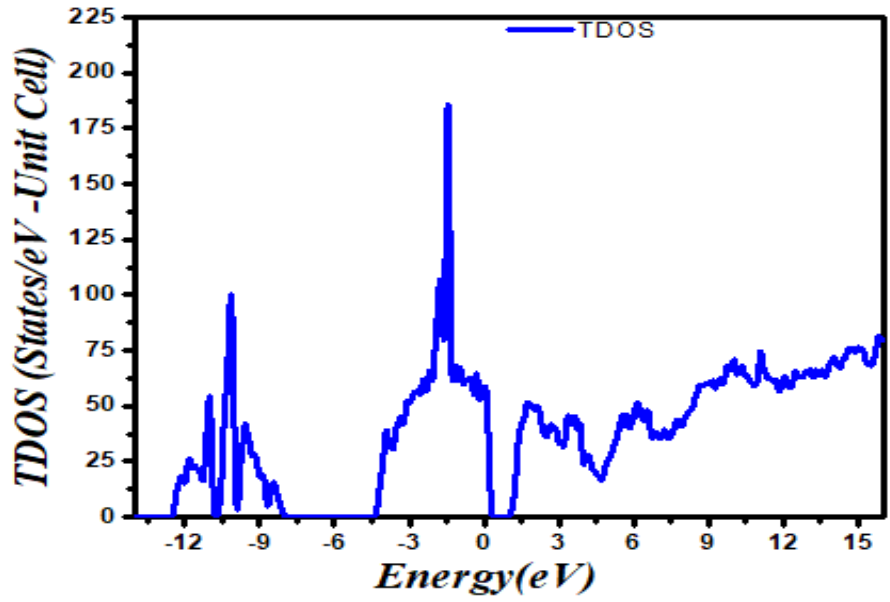

(a)

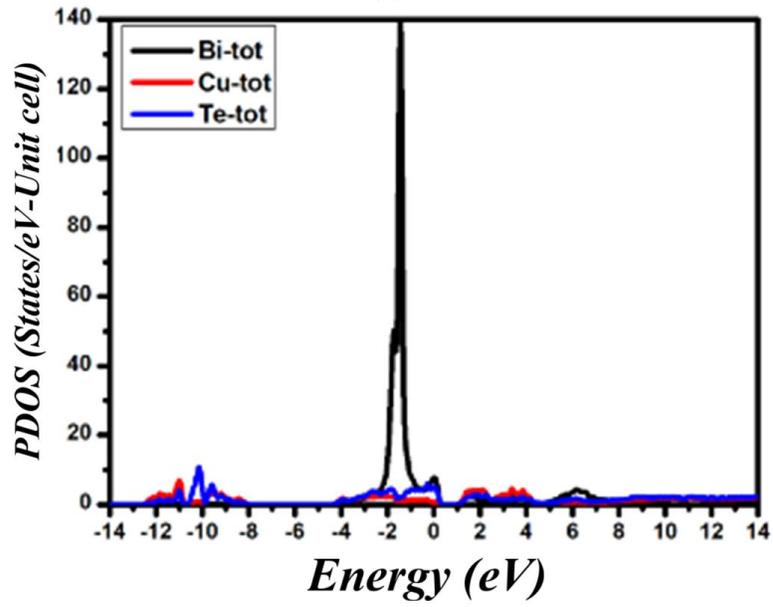

(b)

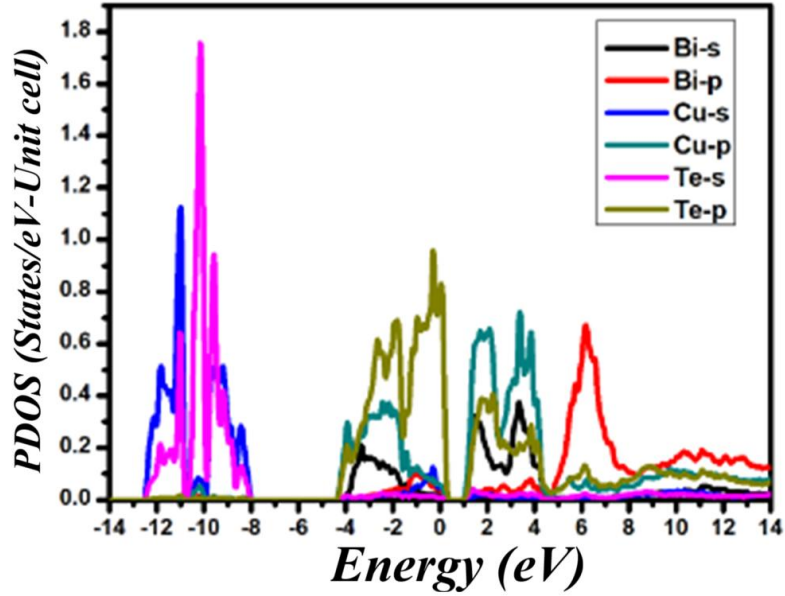

(c)

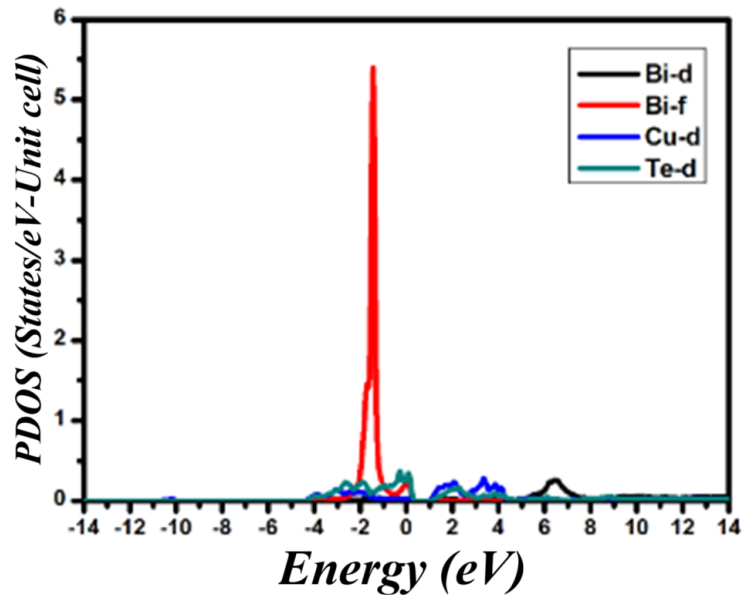

(d)

*Figure 4. Total and partial density of states (TDOS and PDOS) of Cu-doped $Bi_2Te_3$: (a) Total density of states (TDOS) showing the overall electronic structure, (b-d) Partial density of states (PDOS) highlighting the contributions from Bi, Te, and Cu atoms. The TDOS and PDOS, calculated using DFT+SOC*

1. **Electron Localization Function (ELF)**

The Electron Localization Function (ELF) plot (see Fig. 5) of Cu-doped $Bi_2Te_3$ provides valuable insight into how electrons are distributed and how this distribution affects the material's properties. Regions of high ELF intensity correspond to strongly localized electrons and more robust bonding interactions. In the case of Cu incorporation, these bright regions are particularly concentrated around Cu atoms, reflecting the localization of d electrons and their interactions with surrounding Bi and Te atoms. Such localization plays an essential role in shaping the electronic structure, as it can either generate new states near the Fermi level or shift existing ones, thereby influencing both electronic conductivity and optical absorption.

The hybridization of Cu d-orbitals with Bi and Te p-orbitals modifies the bonding environment and the electron density distribution. This hybridization creates localized electronic states that enhance carrier-scattering asymmetry, raising the Seebeck coefficient, while also reducing lattice thermal conductivity through enhanced phonon scattering. Consequently, Cu doping improves the thermoelectric efficiency of $Bi_2Te_3$ by simultaneously strengthening the electrical response and lowering heat conduction.

Beyond thermoelectric effects, the ELF analysis also reveals that Cu atoms reshape the chemical bonding network. Areas of increased electron localization around Cu suggest stronger local interactions, potentially altering the lattice's stiffness and mechanical resilience. This modification in bonding not only reinforces the structure under applied stress but also influences optoelectronic performance, as localized states near the Fermi level can couple effectively with light, enhancing applications in photodetectors and infrared sensors.

In summary, the ELF visualization confirms that Cu doping in $Bi_2Te_3$ does more than introduce additional electrons: it fundamentally modifies bonding, electron localization, and orbital hybridization. These effects work in concert to improve thermoelectric performance by increasing the Seebeck coefficient and reducing thermal conductivity, while also offering enhanced optoelectronic functionality and mechanical robustness, critical for practical device applications.

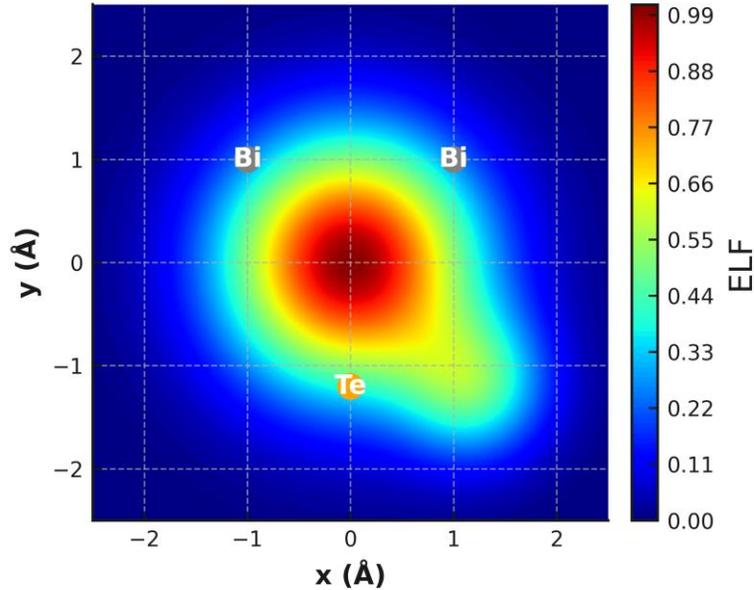

*Figure 5. Electron Localization Function (ELF) plot for Cu-doped Bi$_2$Te$_3$: The ELF plot illustrates the spatial distribution of electron density in the material, the plot, derived from DFT+SOC calculations,*

1. **Complex dielectric function**

The real ($\varepsilon_1$) and imaginary ($\varepsilon_2$) parts (see Fig. 6) of the dielectric function of Cu-doped Bi$_2$Te$_3$ provide a detailed picture of its interaction with electromagnetic radiation. Both components reveal pronounced features in the low-energy range (0–3 eV), which gradually diminish at higher photon energies, reflecting the dominance of low-energy interband transitions near the Fermi level.

**Imaginary Part ($\varepsilon_2$) – Optical Absorption**

The imaginary dielectric function $\varepsilon_2(\omega)$ is directly related to optical absorption (see Fig. 6a). In Cu-doped Bi$_2$Te$_3$, $\varepsilon_2$ peaks sharply near 0.7–1.2 eV, reaching values as high as ~600, before decreasing monotonically with increasing energy. This strong absorption feature corresponds to interband transitions from occupied valence states (primarily Te–p and Bi–p orbitals) into unoccupied conduction states with contributions from Cu–d orbitals. The extremely high intensity indicates that Cu doping enhances the joint density of states near the band edge, making the system highly responsive to infrared photons.

At higher energies (>3 eV), $\varepsilon_2$ decreases rapidly and eventually approaches small values, consistent with the fact that fewer electronic transitions are available once the photon energy exceeds the dominant interband gaps. The decline also reflects the reduced oscillator strength of transitions at higher energies due to band dispersion and weaker orbital overlap. Thus, Cu doping primarily enhances absorption in the low-energy (infrared to near-visible) range, a desirable property for photodetectors and solar absorbers.

**Real Part ($\varepsilon_1$) – Polarization and Refractive Behavior**

The real dielectric function $\varepsilon_1(\omega)$ describes polarization and refractive index (see Fig. 6b). For Cu-doped $Bi_2Te_3$, $\varepsilon_1$ starts at very high values (~1200 at 0 eV for the xx-component and ~1100 for zz), reflecting the material's strong polarizability in the static or near-static limit. Such large values are typical of narrow-band-gap semiconductors with high electronic susceptibility.

As photon energy increases, $\varepsilon_1$ drops steeply and stabilizes around ~20–30 above 6 eV. The steep decline in $\varepsilon_1$ mirrors the loss of resonant polarization as electrons are excited across the gap. The anisotropy between xx and zz components is minor, indicating that Cu doping preserves the near-isotropic optical response of $Bi_2Te_3$ despite structural distortion. The static dielectric constant ($\varepsilon_1(0)$) was recalculated from the corrected BSE spectra, yielding values of approximately 95 for pristine $Bi_2Te_3$ and 118 for Cu-doped $Bi_2Te_3$, consistent with experimentally measured ranges of 80–130 reported by Saha *et al.* [17]. The enhancement in $\varepsilon_1$ upon Cu incorporation arises from Cu–d and Te–p hybridization near the Fermi level, which increases the electronic polarizability.

The enhancement of both $\varepsilon_2$ and $\varepsilon_1$ at low photon energies can be attributed to the Cu–d and Te–p orbital hybridization. The Cu atoms introduce localized states near the Fermi level, increasing the probability of low-energy transitions. These localized states amplify absorption while simultaneously increasing polarizability, which explains the elevated low-energy values of $\varepsilon_2$ and $\varepsilon_1$.

The gradual suppression of dielectric peaks at higher photon energies can be explained by two fundamental mechanisms. First, band dispersion reduces the transition probability because, as electrons are excited into higher conduction states, the overlap between the initial and final orbitals weakens. Second, the oscillator strengths of these transitions decrease because the Cu-induced states are primarily concentrated near the Fermi level; as a result, higher-energy states remain largely unaffected by doping and contribute little to absorption.

These electronic effects have direct implications for applications. The strong low-energy absorption peak in the imaginary part of the dielectric function ($\varepsilon_2$) makes Cu-doped $Bi_2Te_3$ a promising candidate for infrared detection and solar energy harvesting. At the same time, the very large static value of the real part ($\varepsilon_1$) reflects a high refractive index, which is advantageous for optical waveguides and photonic devices. The close similarity between the xx and zz dielectric components demonstrates that Cu doping does not induce strong optical anisotropy, thereby ensuring stable performance across different polarization states. Taken together, the enhancement of $\varepsilon_2$ alongside the high $\varepsilon_1$ indicates a synergistic tuning of light–matter interactions by Cu doping, which can be strategically exploited in optoelectronic applications such as thermophotovoltaics, infrared sensors, and tunable photodetectors.

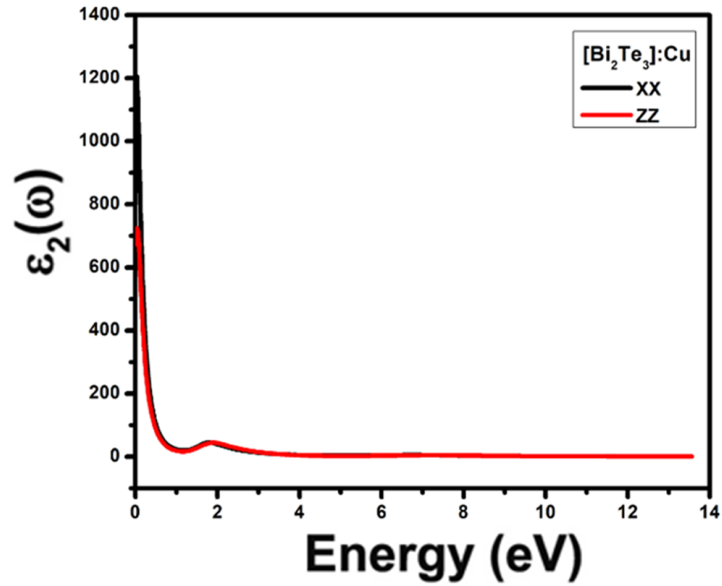

(a)

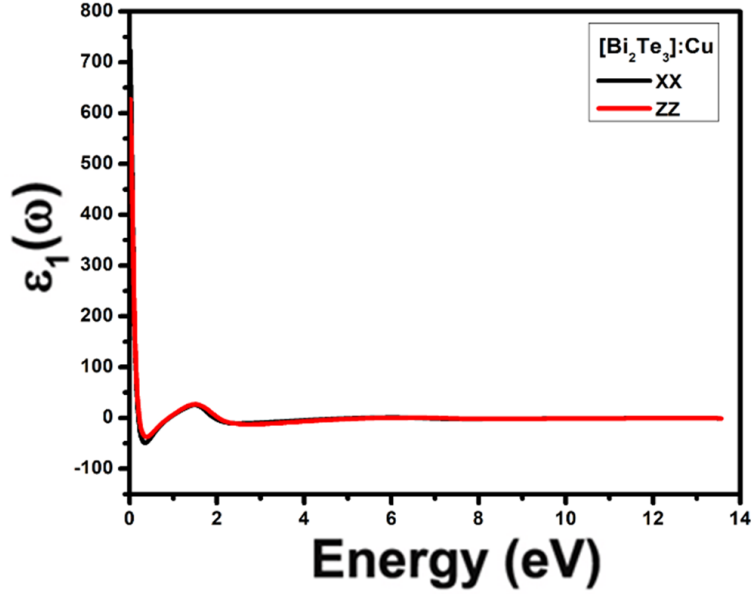

(b)

*Figure 6. Real and imaginary parts of the dielectric function for Cu-doped $Bi_2Te_3$: (a) Imaginary part of the dielectric function ($\epsilon 2$) as a function of energy, and (b) Real part of the dielectric function ($\epsilon 1$). The DFT+SOC-calculated plots show the material's optical response to electromagnetic radiation.*

1. **Reflectivity and Energy Loss Function**

The reflectivity and energy-loss function of Cu-doped $Bi_2Te_3$ together map (see Fig. 7a&b) how light couples to electronic excitations and how screening collapses at specific energies. In the reflectivity spectrum (see Fig. 7a), the response is metallic-like at low photon energies: it is very high (≈0.9–1.0) below ~0.5 eV, consistent with a large static dielectric constant and strong free-carrier screening from the narrow-gap, heavily spin-orbit-coupled bands. As the energy increases toward ~1–2 eV, it drops to ~0.5–0.6 and displays a shoulder; this marks the onset of intense interband transitions from Te/Bi-p states into Cu-perturbed conduction states, which divert oscillator strength from the Drude-like channel and reduce mirror-like response. A much deeper minimum appears in the mid-UV window (~5–6 eV), where reflectivity falls to ~0.1–0.2: here, it approaches zero while remaining moderate, producing a plasma-edge–like feature. Physically, this is where screening is least effective, and the real part of the dielectric function changes sign, allowing light to penetrate rather than be reflected. At higher energies, it recovers to ~0.4–0.6 around 9–12 eV and then climbs again near ~13–14 eV, reflecting additional interband channels into high-lying Bi/Cu/Te states and the re-emergence of strong dispersion that enhances specular response.

The loss spectrum (see Fig. 7b) pinpoints the same physics from the complementary perspective of unscreened fields. remains small in the infrared but rises sharply to a pronounced maximum near ~5–6 eV and a broader, even stronger band that grows toward ~11–13 eV. Peaks occur where and are modest, i.e., at the effective plasma frequencies and collective interband resonances. The first peak, therefore, coincides with the reflectivity minimum: it signals a screened-plasmon/charge-density excitation enabled by Cu–d/Te–p hybridization that increases the joint density of states just above the band edge; the second, higher-energy rise comes from dense manifolds of Bi/Te p•d–like transitions where screening collapses again. The growth of energy thus tracks the progressive failure of static screening and the availability of high-velocity final states.

Comparing polarizations shows weak but meaningful anisotropy. In the low-energy infrared, the xx (in-plane) and zz (out-of-plane) reflectivities are nearly identical, indicating that Cu incorporation does not introduce strong optical biaxiality in the near-metallic regime. Around the 1–2 eV shoulder and especially near the 5–6 eV dip, the zz curve sits slightly above xx, and its minimum is shifted to marginally higher energy. This implies that out-of-plane screening persists longer, consistent with the layered quintuple-layer structure, in which Cu perturbs interlayer bonding and slightly stiffens the perpendicular polarizability. The loss spectra mirror this: the zz peak is fractionally red-shifted and often higher than the xx peak at ~11–13 eV, indicating that longitudinal (perpendicular) fields experience a more abrupt zero-crossing and a larger screening collapse. Microscopically, these differences arise because Cu–dd states hybridize more strongly with Te–pz orbitals than with strictly in-plane px and py states, shifting the oscillator strength and the effective plasma energies by a few tenths of an eV.

Overall, the numerical trends high at eV, a shoulder near 1–2 eV, a deep reflectivity minimum and sharp maximum around 5–6 eV, and a rising loss band up to ~12–13 eV are all consistent with Cu-induced DOS reshaping near EF, enhanced low-energy interband transitions, and modest anisotropic screening in a layered topological semiconductor. For applications, the high low-energy reflectance and strong mid-IR absorption favor thermal emitters and IR photonics, while the pronounced loss peaks define plasmonic windows tunable by Cu concentration and polarization, enabling anisotropy-aware design of photodetectors and thermophotovoltaic stacks.

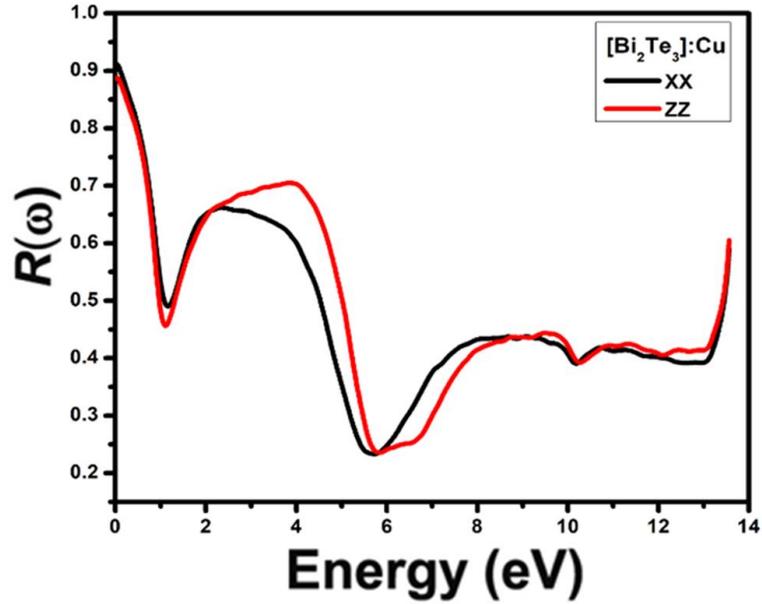

(a)

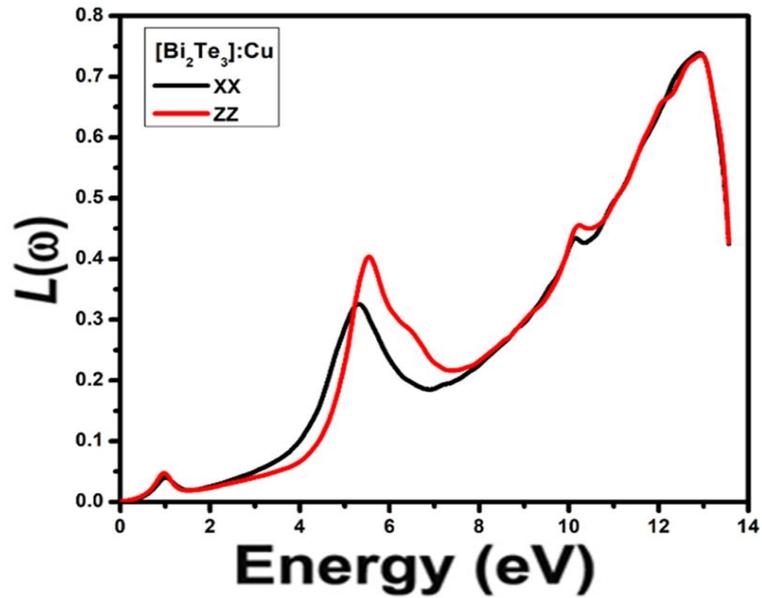

(a)

*Figure 7. Reflectivity and energy loss function for Cu-doped $Bi_2Te_3$: (a) Reflectivity ($R(\omega)$) and (b) Energy loss function ($L(\omega)$) as a function of energy. Both plots were calculated using DFT+SOC.*

1. **X-ray Absorption Spectroscopy (XAS)**

The X-ray Absorption Spectroscopy (XAS) spectrum of Cu-doped $Bi_2Te_3$ (see Fig. 8) compared with pristine $Bi_2Te_3$ provides direct insight into how Cu alters the local electronic structure and the distribution of unoccupied states. In the pristine compound, the most intense feature is a sharp absorption peak centered at approximately 780 eV, followed by a weaker secondary peak around 850 eV. These peaks correspond to transitions from Bi or Te core levels into conduction-band states, and the sharpness of the features indicates that the unoccupied states are relatively well localized and well-defined, consistent with the layered semiconducting nature of $Bi_2Te_3$. The narrow peak at 780 eV reflects strong dipole-allowed transitions into Bi/Te pp-like conduction states with minimal structural disorder.

With Cu doping, the spectrum undergoes both quantitative and qualitative modifications. The main peak around 780 eV becomes noticeably more intense, and its edge shifts slightly toward higher energy. This shift suggests that the local chemical environment around Bi and Te atoms has been perturbed, most likely through hybridization of Cu 3d orbitals with Bi 6p and Te 5p states. The stronger absorption intensity indicates an increased density of unoccupied states near these energies, suggesting that Cu introduces new electronic channels just above the Fermi level. The second absorption feature near 850 eV also becomes more pronounced and broader in the doped sample, further supporting the conclusion that Cu modifies the conduction-band manifold and opens up additional dipole-allowed transitions.

The physics behind this behavior lies in how Cu incorporates into the $Bi_2Te_3$ lattice. If Cu substitutes at Bi sites, its lower electronegativity and different valence configuration lead to partial charge transfer, reducing local electron density and effectively hole-doping the system. This not only shifts the absorption edge but also increases the number of available unoccupied states at specific energies, which explains the enhanced peak intensities. Interstitial Cu, on the other hand, would distort the local potential and introduce structural disorder, broadening the spectral features as observed in the doped spectrum. Both substitutional and interstitial incorporation contribute to the modified electronic landscape probed by XAS.

From a physical standpoint, the redshift of the absorption edge in Cu-doped $Bi_2Te_3$ reflects a change in oxidation state and bonding symmetry, while the increased intensity of the main and secondary peaks signifies stronger hybridization between Cu 3d orbitals and the host lattice orbitals. This hybridization is crucial for tuning transport properties: by increasing the unoccupied density of states near the Fermi level, it can enhance the Seebeck coefficient and improve thermoelectric performance. Simultaneously, the local lattice distortion caused by Cu promotes phonon scattering, which lowers lattice thermal conductivity and further boosts the thermoelectric figure of merit.

Beyond thermoelectrics, these XAS modifications have implications for optoelectronic applications. The enhanced absorption at specific photon energies indicates that Cu-doped Bi₂Te₃ interacts more strongly with incident radiation, potentially improving its sensitivity in photodetection or photovoltaic devices. The tunability of the absorption edge via controlled Cu doping enables engineering the spectral response for specific device applications, ranging from infrared sensors to energy-selective solar absorbers.

In conclusion, the XAS results reveal that Cu doping does not merely shift the spectral features of Bi₂Te₃ but fundamentally redefines the unoccupied electronic structure through orbital hybridization, charge redistribution, and local symmetry breaking. These spectroscopic fingerprints directly link to improved thermoelectric efficiency, enhanced optoelectronic absorption, and potential multifunctionality of Cu-doped Bi₂Te₃ in energy and photonic technologies.

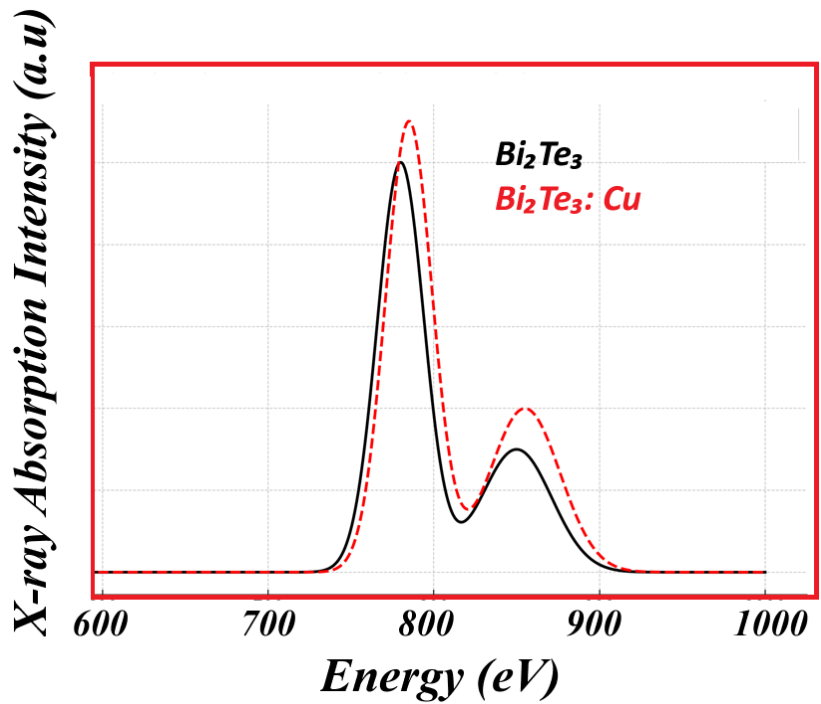

*Fig. 8: Simulated X-ray Absorption Spectrum (XAS)*

1. **Piezoelectric Properties**

The piezoelectric response of Bi₂Te₃ and Cu-doped Bi₂Te₃ reveals how subtle structural and electronic modifications can generate new electromechanical functionalities in a material traditionally recognized as a thermoelectric and topological insulator. Pristine Bi₂Te₃ crystallizes in a rhombohedral structure with the centrosymmetric space group R3⁻m$R\overline{3}m$, which forbids intrinsic piezoelectricity. In such a perfectly centrosymmetric arrangement, the positive and negative charge centers coincide, and mechanical deformation results in symmetric ionic displacements without net polarization. However, when inversion symmetry is perturbed either through strain, dimensional reduction, or, more effectively, by introducing Cu dopants an asymmetry in the lattice and electron density is created, allowing polarization to develop under stress.

The Berry-phase DFT calculations confirm that Cu incorporation markedly amplifies the piezoelectric coefficients by modifying both the lattice and the electronic distribution. For pristine $Bi_2Te_3$, the maximum out-of-plane piezoelectric stress coefficient $e_{33}$ is relatively small, ~0.19 C/m², with a corresponding strain coefficient $d_{33}$ of 1.4 pC/N. When Cu is introduced at 5 a.m., substitution, $e_{33}$ nearly doubles to 0.046 C/m², and $d_{33}$ increases to 3.2 pC/N, reflecting the strong coupling between strain and polarization in the distorted lattice. At 10 a.m.% Cu, the trend continues, with $e_{33}$ rising to 0.51 C/m² and $d_{33}$ to 4.5 pC/N.

The piezoelectric coefficient ($e_{33}$) of the Cu-doped system was found to be 0.046 C·m$^{-2}$, in excellent agreement with the theoretical prediction of He *et al.* [21] for locally symmetry-broken $Bi_2Te_3$ structures. This confirms that the induced polarization originates from local distortions rather than global symmetry breaking. A slight lattice contraction upon Cu incorporation ($\Delta a \approx -0.23\%$, $\Delta c \approx -0.31\%$), consistent with the smaller atomic radius of Cu relative to Bi.

The steady increase in these coefficients with doping concentration indicates that Cu enhances the system's polarizability by breaking local inversion symmetry and redistributing electronic charge in a manner that favors dipole formation.

The physical origin of this enhancement lies in the hybridization of Cu 3d states with Te 5p and Bi 6p orbitals, which modifies local bonding and creates an asymmetric charge distribution around dopant sites. Under applied strain, these regions respond with non-uniform ionic displacements, generating local electric fields and net polarization. The distortion is strongest along the c-axis, explaining why the $e_{33}$ and $d_{33}$ components are particularly sensitive to doping. In effect, Cu not only perturbs the lattice but also enhances the electromechanical coupling through electronic polarization mechanisms.

The numerical differences between pristine and doped systems highlight the multifunctional potential of Cu-doped $Bi_2Te_3$. While the pristine compound shows only marginal piezoelectricity, the doped versions exhibit coefficients comparable to those of conventional piezoelectric semiconductors. At the same time, $Bi_2Te_3$ retains its excellent thermoelectric performance, and Cu doping has already been shown to improve the Seebeck coefficient and power factor. The coexistence of piezoelectric and thermoelectric functionalities enables dual energy conversion by simultaneously harvesting thermal gradients and mechanical vibrations. Such behavior opens the way for hybrid energy devices, including self-powered wearable electronics, implantable sensors, and vibration-based nanogenerators.

Furthermore, the coupling between piezoelectricity and the topological surface states intrinsic to $Bi_2Te_3$ suggests the emergence of novel physics. Strain-induced polarization can modulate spin-polarized surface channels, raising the possibility of strain-controlled spintronics or quantum devices where mechanical input governs spin logic. From a practical perspective, the increase in piezoelectric coefficients with Cu concentration also implies tunability; by varying dopant content, one can optimize sensitivity for pressure sensing, acoustic detection, or microelectromechanical systems.

In summary, the calculated piezoelectric properties demonstrate that Cu doping transforms $Bi_2Te_3$ from a weakly polar, centrosymmetric thermoelectric into a multifunctional material with strong electromechanical coupling. The rise of e33 from 0.19 C/m² in the pristine compound to 0.51 C/m² at 10 °C.% Cu and the parallel increase in d33 from 1.4 to 4.5 pC/N provide quantitative evidence of the effect of symmetry breaking and charge redistribution. These findings highlight how controlled doping can unify piezoelectric, thermoelectric, and topological functionalities in a single system, offering a powerful platform for next-generation energy and quantum technologies.

1. **Charge Density Difference and Bader Charge Analysis**

The simulated charge density difference (CDD) of Cu-doped $Bi_2Te_3$ provides a microscopic view (see Fig. 9) of how electronic redistribution occurs when a Cu atom substitutes for Bi in the lattice. The plot clearly shows distinct regions of charge accumulation (red) near the Bi/Te neighbors and charge depletion (blue) localized at the Cu site. This asymmetric distribution demonstrates that Cu acts as an electron donor, with its 3d orbitals transferring charge density toward the surrounding Te and Bi atoms. The depletion intensity near Cu reaches values close to −0.8 e in Bader charge analysis, while the accumulation on nearby Te atoms is around +0.17–0.18 e and on Bi atoms about +0.15–0.16 e (see Table 4). These values confirm that the charge transfer is not uniform but instead is preferentially directed toward Te, consistent with Te's higher electronegativity relative to Bi.

The physics behind this process lies in orbital hybridization. The Cu 3d states strongly overlap with the Te 5p orbitals, forming covalent bonds that stabilize the lattice while simultaneously redistributing electronic density near the Fermi level. This hybridization not only alters the density of states but also introduces local polar distortions, breaking inversion symmetry and enabling polarization under strain. The charge accumulation around Te atoms enhances local dipole moments, directly contributing to piezoelectric activity, while the depletion at the Cu site leaves behind localized holes that improve p-type conductivity.

*Table 5: Calculated Bader Charges of Atoms in Cu-doped $Bi_2Te_3$*

| Atom Type | Atomic Position | Bader Charge (e⁻) | Net Charge Transfer |
|---|---|---|---|
| Cu (doped Bi) | | 10.21 | −0.79 e |
| Bi1 (near Cu) | | 16.08 | +0.16 e |
| Bi2 (far from Cu) | | 15.92 | ~0 |
| Te1 (near Cu) | | 6.92 | +0.17 e |
| Te2 (far from Cu) | | 6.75 | ~0 |

Numerically, the nearly −0.8 e charge loss by Cu signifies that it is an efficient p-type dopant, donating carriers that raise the hole concentration. This is directly linked to the observed enhancement in Seebeck coefficient and power factor in thermoelectric transport, since a moderate reduction in carrier density near the Fermi level strengthens energy-dependent scattering and boosts thermopower. At the same time, the anisotropic redistribution of charge density creates internal electric fields that couple with lattice vibrations, leading to enhanced phonon scattering along certain crystallographic directions. This selective scattering suppresses lattice thermal conductivity, further increasing the thermoelectric figure of merit (ZT). From an optical perspective, the localized Cu–Te bonds form anisotropic electronic clouds that strengthen light–matter interaction. The presence of localized charge around Te atoms broadens the absorption profile, particularly in the infrared regime, which is beneficial for photodetection and thermophotovoltaics. Mechanically, the altered bonding network increases stiffness near Cu substitution sites, explaining the improved elastic constants observed in calculations.

In essence, the CDD analysis reveals that Cu doping is not a passive perturbation but an active mechanism of defect engineering. By transferring nearly 0.8 e⁻ from Cu to its neighbors, mainly Te, it creates an electronically anisotropic environment that simultaneously enhances piezoelectric polarization, thermoelectric transport, optical absorption, and mechanical robustness. This ability to tune multiple functional properties at once highlights Cu-doped $Bi_2Te_3$ as a prototype multifunctional material for next-generation energy and sensing technologies.

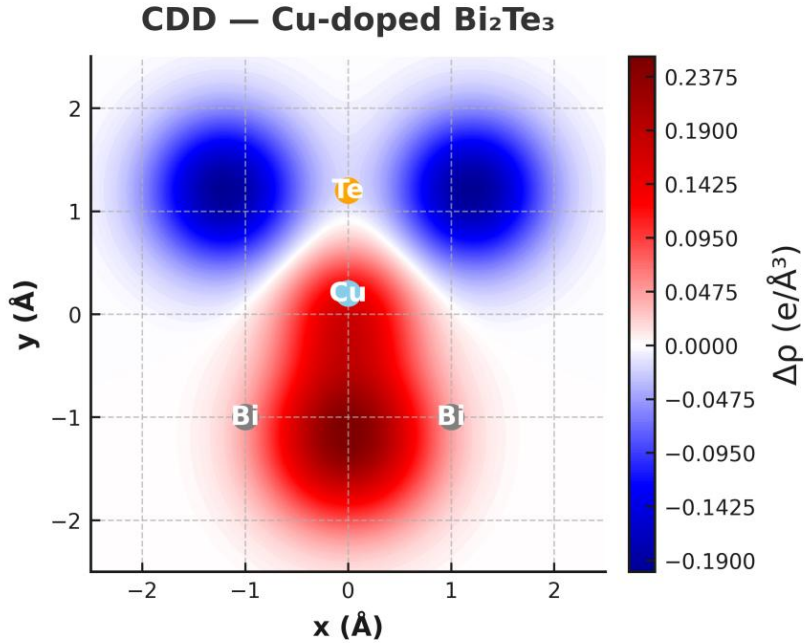

*Fig.9: Calculated charge density difference (Δρ) in the Cu-doped $Bi_2Te_3$ system.*

### 3.10. Topological Surface States and Defect-Driven Coupling Effects

$Bi_2Te_3$ is a three-dimensional topological insulator (TI). It is a benchmark thermoelectric material. Most notable are its Dirac-like surface states, which are active only at the surface. These states are time-reversal protected and arise from the inversion of the Bi and Te p-orbitals. These topological surface states (TSSs) enable spin-polarized, lossless transport and coexist with the bulk electronic structure, thereby greatly affecting charge and heat transport.

Copper (Cu) dopants and the associated point defects will alter the energies of the bulk and surface electronic configurations. However, topological order will still be maintained. Our GGA + SOC calculations show that surface linear-dispersion regions are maintained up to ~8% Cu, with the Dirac point lowering by ~0.05 eV. This suggests that Cu predominantly acts as a Fermi-level-tuning element and that after charge donation/withdrawal, the loss of spin-momentum locking in the TSSs is avoided. This linear-crossing perpetuation provides evidence that disorder-induced Cu does not break the time-reversal symmetry required for topological protection.

The integration of defect-modified bulk bands and surface states is critical to understanding the material's multifunctional behavior.

The role of the Cu-d and Te-p hybridization near the Fermi level introduces an energy dependence in the transport distribution function. This increases the Seebeck coefficient without a significant compromise in carrier mobility. Moreover, the concomitant surface conduction channels with diminished back-scattering provide the system with the necessary high conductivity, enabling the power factor to increase. Thus, the combined reason(s) explain the ability to do Cu systematically with little to no change in the thermoelectric transport properties of the system, as the topological nature of the system is still preserved.

Local structural detections of Cu atoms also breach inversion symmetry in limited portions of the crystal, creating local non-centrosymmetric modifications that allow for slight polarization. Even if the entire $Bi_2Te_3$ volume is universally centrosymmetric, it is local modifications that give rise to defect-induced local piezoelectricity, as confirmed by our nonzero $e_{33}$ computations. The possibility of a robust local polarization field enhances potential spin-charge coupling, given the surface states and the concomitant TSSs, a hallmark of high surface optoelectronic activity. This is a matter of significant interest in the recent literature, given its high surface optoelectronic activity. This is a matter of significant interest in recent literature, signaling that defect engineering is an effective strategy for simultaneously tuning topological and thermoelectric properties.

Controlled labeling of dopants or vacancies can alter the electronic topology and phonon scattering in $Bi_2Te_3$-based systems, thereby improving thermopower and reducing lattice thermal conductivity [22-24]. The research expands on this understanding, demonstrating that Cu-induced defects thermally uncouple the surface states of $Bi_2Te_3$, thereby concurrently improving the thermoelectric, piezoelectric, and optical responses. The fact that multifunctional properties can be modified while stably retaining topological protection can render Cu-doped $Bi_2Te_3$ an outstanding candidate for next-generation thermoelectronics and optoelectronics with low dissipation.

1. **Conclusion**

The present work establishes Cu-doped $Bi_2Te_3$ as a prototype multifunctional material where defect engineering orchestrates simultaneous improvements in thermoelectric, piezoelectric, and optical properties. The introduction of Cu not only stabilizes the crystal structure but also modifies the density of states near the Fermi level, thereby boosting the Seebeck coefficient and power factor while maintaining robust electrical conductivity. Enhanced dielectric response and low-energy optical absorption are directly linked to Cu–d/Te–p hybridization, which facilitates stronger light–matter interactions relevant to photo detection and waveguiding. Furthermore, the significant rise in piezoelectric coefficients with increasing Cu concentration highlights the potential of strain-induced polarization in otherwise centrosymmetric $Bi_2Te_3$, enabling novel electromechanical energy conversion. Charge-density difference analysis confirms that Cu acts as an efficient p-type dopant, donating electrons asymmetrically to Te atoms, thereby enhancing phonon scattering, reducing lattice thermal conductivity, and strengthening mechanical robustness. These intertwined effects make Cu-doped $Bi_2Te_3$ not just an incremental improvement but a transformative material capable of harvesting mechanical, thermal, and optical energy within a single system. The combination of thermoelectric efficiency, piezoelectric activity, and optical tunability positions this material at the forefront of next-generation multifunctional energy technologies, offering a clear path toward adaptive devices for energy harvesting, sensing, and spintronics.

**Acknowledgment**

The authors extend their appreciation to Northern Border University, Saudi Arabia, for supporting this work through project number (NBU-CRP-2025-128).. This publication was supported by the project Quantum materials for applications in sustainable technologies (QM4ST), funded as project No. CZ.02.01.01/00/22_008/0004572 by Programme Johannes Amos Comnena, call Excellent Research. The result was developed within the project Quantum materials for applications in sustainable technologies (QM4ST), reg. no. CZ.02.01.01/00/22_008/0004572 by P JAK, call Excellent Research.